# Unsupervised Deep Image Prior for Sparse-View and Limited-Angle Electron Tomography

Serge Brosset [a], Daniel del Pozo Bueno [a], Thomas David [b], Laure Guetaz [b], Philippe Ciuciu [c,d] and Zineb Saghi [a,*]

[a] Univ. Grenoble Alpes, CEA, Leti, F-38000 Grenoble, France.
[b] Univ. Grenoble Alpes, CEA, Liten, F-38000 Grenoble, France.
[c] CEA, Joliot, NeuroSpin, Gif-sur-Yvette Cedex 91191, France.
[d] Inria, MIND, Université Paris-Saclay, Palaiseau 91120, France.

**E-mail address:**

Serge.brosset@proton.me (Serge Brosset), daniel.delpozobueno@cea.fr (Daniel Del Pozo Bueno), thomas.david3@cea.fr (Thomas David), laure.guetaz@cea.fr (Laure Guetaz), philippe.ciuciu@cea.fr (Philippe Ciuciu), zineb.saghi@cea.fr (Zineb Saghi).

*** Corresponding author:**

Zineb Saghi
Commissariat à l'énergie atomique et aux énergies alternatives
MINATEC Campus
17 rue des martyrs
F-38054
Grenoble Cedex 9.
Tel: +33 (4) 38 78 48 70
Email: zineb.saghi@cea.fr

## Abstract

Electron tomography (ET) plays an important role in the three-dimensional (3D) characterization of nanomaterials. However, under limited-angle and sparse-view conditions, conventional algorithms produce degraded reconstructions, which compromise the quality and interpretability of resulting 3D data. In this paper, we present deep image prior (DIP), an unsupervised deep learning (DL) approach, for highly degraded tomography acquisitions and demonstrate, using simulated data, that its performance is comparable to that of supervised approaches requiring training datasets, even for tilt ranges as limited as 60° and tilt increments of 10°. We then apply it to experimental data and show that it enables reliable 3D quantification under both sparse-view and limited-angle conditions, highlighting its potential for a wide range of materials and acquisition modalities.

## Introduction

Electron tomography (ET) based on scanning transmission electron microscopy (STEM) is a powerful technique for elucidating and optimizing material properties. Using state-of-the-art microscopes, ET enables three-dimensional (3D) characterization of the morphology, structure and chemistry of nanomaterials at (sub-)nanometer spatial resolution [1]. Ideally, a tomography experiment involves acquiring projections over a complete tilt range (±90°) with fine tilt increments of 1-2°. Under these conditions, conventional reconstruction algorithms, such as filtered back projection (FBP) and the simultaneous iterative reconstruction technique (SIRT), yield high-quality reconstructions with isotropic resolution, thereby facilitating accurate segmentation and interpretation of the resulting 3D reconstructions [2-3].

In practice, most TEM samples are prepared on grids or as thin lamellae. Due to this slab-like geometry, even when using a dedicated tomography holder, the accessible tilt range is typically limited to approximately ±80°. This restriction arises from spatial constraints within the pole-piece gap of the TEM objective lens and shadowing effects caused by the specimen holder. Increased sample thickness and overlapping structures outside the imaged region at high tilt angles further reduce the achievable tilt range, often to about ±60° [4-5]. Additional challenges emerge when ET is conducted *in situ*, as such experiments require dedicated holders that are generally thicker than conventional ET holders, thereby further restricting the accessible tilt range ±50° or less [6-7]. Limited-angle acquisitions result in a missing-wedge (MW) of information in the objects' Fourier space. Consequently, reconstructions using FBP and SIRT exhibit the so-called *missing-wedge artifacts*, characterized by blurring and elongation of features along the beam direction. These artifacts lead to anisotropic resolution and reduce the accuracy and reliability of 3D segmentation.

More advanced reconstruction algorithms, such as compressed sensing (CS) methods, have been shown to produce high-quality reconstructions from undersampled tomography data [8]. CS methods involve adding a regularization term to the data fidelity term to promote sparsity of the object in a chosen basis (e.g.: the gradient, the wavelet transform, or a combination of various transforms). These methods generate reliable reconstructions from sparse-view acquisitions spanning the full tilt range (e.g. [-90°:10°:+90°]) and from datasets with a mild MW region (e.g. [-70°:+70°] with 2° to 5° tilt increments) [9-12]. However, as reported in [13-15], CS methods do not compensate for severe MW artifacts. Similar conclusions were observed with other reconstruction approaches, including the model-based iterative reconstruction (MBIR) method [16-17], the direct iterative reconstruction of computed tomography trajectories approach (DIRECTT) [18], and the discrete algebraic reconstruction technique (DART) [19-20].

In recent years, deep learning (DL) has emerged as a revolutionary framework, outperforming state-of-the-art techniques in image and signal processing. In the field of electron microscopy, DL is increasingly employed for tasks such as image denoising, super-resolution, segmentation, and object detection [21-22]; automated TEM alignment [23]; peak detection and quantification in spectroscopy [24]; and gradient descent optimization in electron ptychography [25].

In the specific context of ET, DL approaches can be applied at different stages in the reconstruction pipeline, as summarized in Figure 1. In [26], the authors reconstructed a 4 nm diameter Pt nanoparticle from a tilt series of 21 projections spanning the angular range [-71.6°:71.6°] and successfully corrected the MW artifacts by training a DL network on simulated tomograms using the atomicity constraint. IsoNet [27] and DeepDeWedge [28] are increasingly used in cryo-ET, with IsoNet enabling isotropic 3D reconstruction and DeepDeWedge performing simultaneous denoising and MW correction. Despite their self-supervised nature, both approaches have two key limitations: they require sub-tomograms containing similar structural features, and they have been demonstrated only under highly sampled, mildly limited-angle conditions ([-60°:2°:+60°]). Finally, we cite [29], where two DL approaches were used for STEM-EDX tomography of nanocrystals under sparse-view and low-dose conditions. Simultaneous STEM-HAADF and STEM-EDX datasets were acquired from -60° to +60°, with a 10° increment. Denoising of the chemical maps was performed using an unsupervised cycleGAN approach, assuming that the average signal of the EDX maps matches the HAADF measurement. The MW was corrected post-reconstruction using an unsupervised ProjectionGAN approach, which implicitly assumes object symmetry. In all the cited papers,

MW artifacts were mitigated *post-reconstruction* (blue path in Figure 1), either using supervised models trained on simulated data or unsupervised ones that impose strong assumptions. In all cases, MW larger than 60° were *not* addressed. Complementary to these *post-reconstruction* approaches, we also cite *pre-reconstruction* strategies (green path in Figure 1), where the DL models are used for inpainting the sinogram [30]; however, these are not addressed in the present work.

Recently, unsupervised approaches have been developed that eliminate the need for training data and strong assumptions. In particular, the deep image prior (DIP) framework has demonstrated remarkable proficiency in tasks such as image restoration, denoising, inpainting, and super-resolution [31]. DIP has also been proposed as a strategy for simultaneous reconstruction and restoration in X-ray tomography [32] (red path in Figure 1). This approach employs an untrained convolutional neural network as an implicit prior and optimizes the network parameters directly on low-quality or degraded data to produce a high-quality 3D volume.

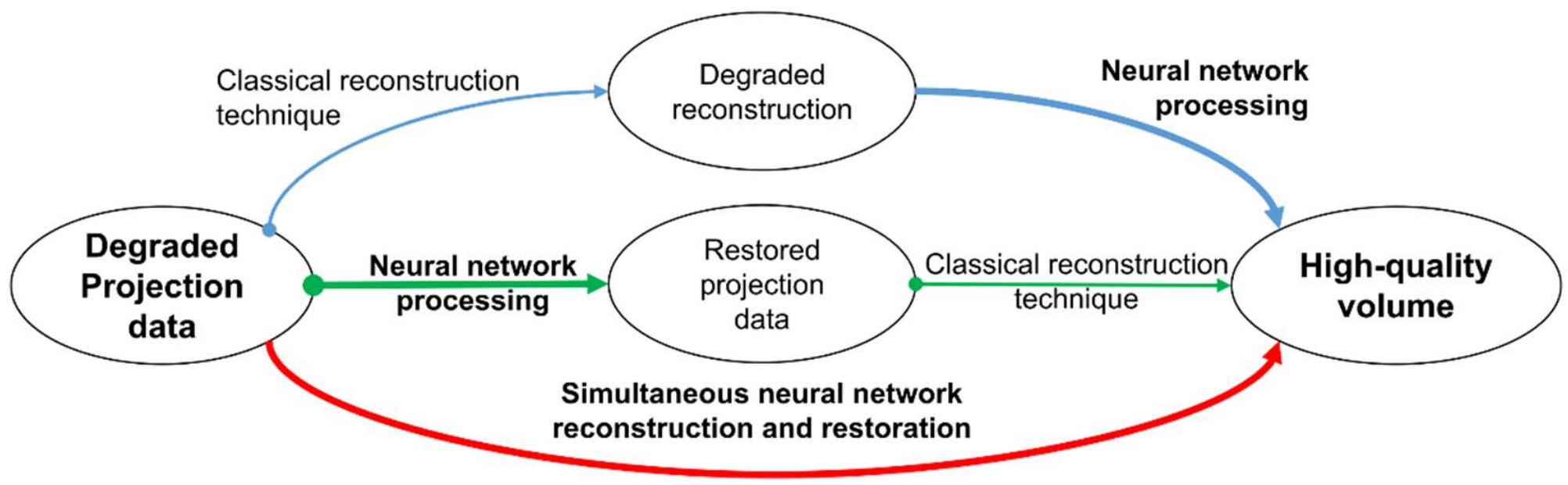

Figure 1: Three DL strategies for limited-angle tomography: post-reconstruction image restoration (blue), pre-reconstruction projection restoration (green), and simultaneous reconstruction and restoration (red).

In this work, we evaluate the robustness of a supervised post-reconstruction approach using a U-Net network and an unsupervised DIP approach across three acquisition scenarios: [-60°:2°:60°], [-60°:10°:60°] and [-30°:2°:30°]. The first acquisition scenario is commonly used in 3D morphological analyses of nanostructures. The second scenario is mostly encountered in spectroscopic ET, using electron energy loss spectroscopy (EELS) and energy dispersive X-ray spectroscopy (EDX) techniques. In this case, as in [30], large tilt increments are chosen as a compromise to generate high-quality spectrum images –requiring a higher electron dose and longer acquisition times than in imaging mode– while minimizing sample damage. The third scenario represents an extreme case, encountered with highly dense and planar samples, or during experiments using holders thicker than dedicated tomography holders (e.g., standard or *in situ* holders).

In addition, we compare both DL approaches with classical methods. First, using simulated data, we show that the evaluated DL methods achieve superior performance across all three acquisition scenarios. Second, we benchmark these approaches on the 3D quantitative analysis of carbon-supported platinum (Pt) nanoparticles (NPs), which serve as catalysts in the electrodes of proton exchange membrane fuel cells. Our results indicate that DL approaches outperform classical methods and provide reliable information on particle shape and size even when the tilt range is severely limited.

On both simulated and experimental data, the DIP approach yields results comparable to those of the supervised U-Net, but without requiring training data or imposing any assumptions, making it applicable to a wide range of materials and acquisition conditions.

A Python library implementing both DL approaches is available on GitHub.

# Methods

**Simulated data generation:**

In order to benchmark the different methods, and for training the supervised U-Net approach, simulated images containing particles of different shapes (mostly spherical) and sizes were created using scikit-learn library (Figure 2). Next, ASTRA Toolbox [33] was used to generate the projections of the simulated tilt series for the different acquisition scenarios. Quantitative analysis and image quality metrics were extracted, using the SIRT reconstruction from [-90°:1°:90°] tilt series as a reference image.

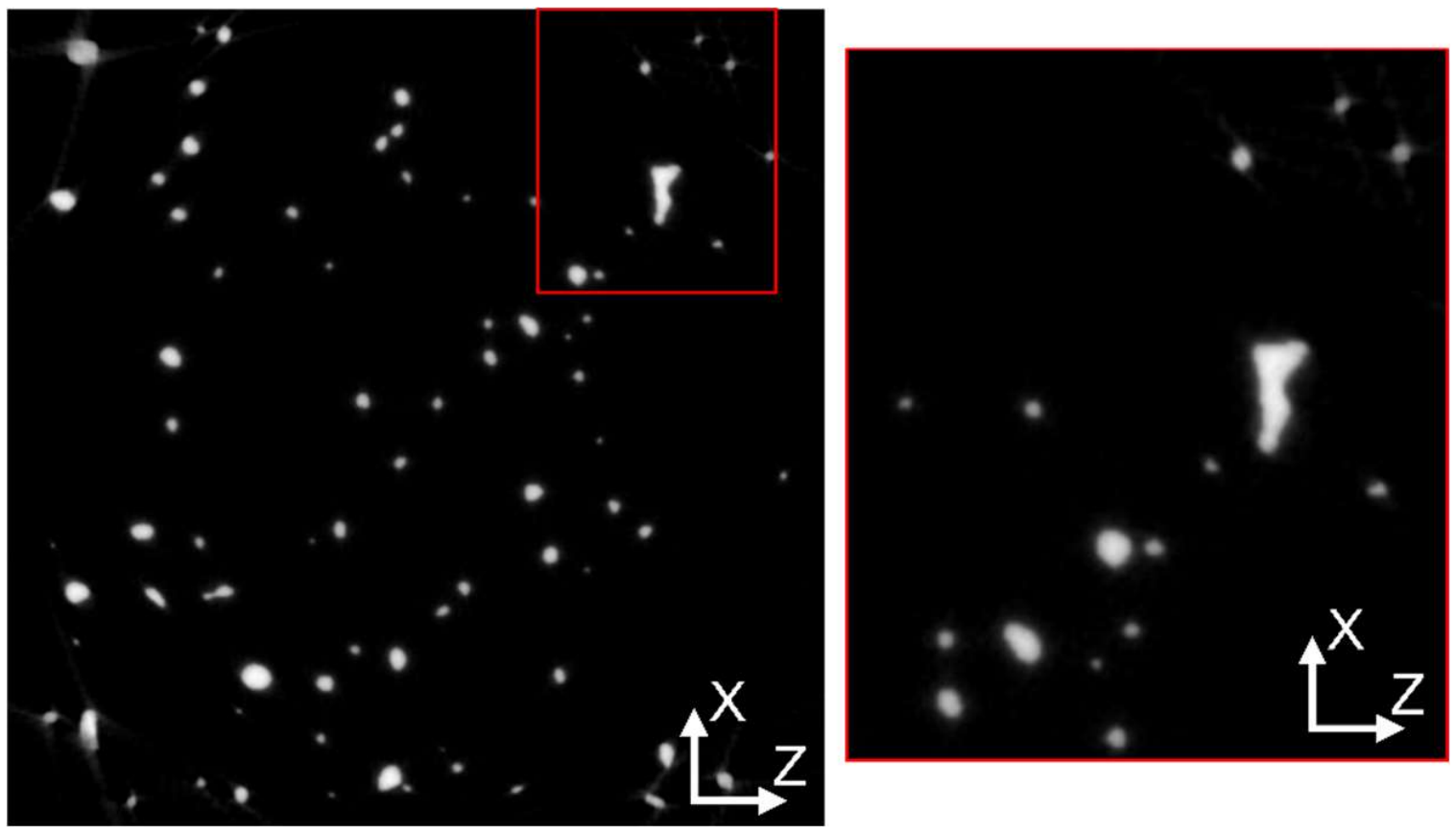

Figure 2: A simulated image containing particles of various shapes and sizes.

**Experimental data and acquisition details:**

Carbon-supported Pt NPs were dispersed onto a carbon grid and mounted on a Fischione 2020 single tilt tomography holder. An HAADF-STEM tilt series was acquired from -60° to +60° with a 2° increment, using a probe-corrected Titan Themis (Thermo Fisher Scientific) operating at 200kV, with a convergence angle of 12mrad, a frame size of 2048x2048 pixels, and a pixel size of 72.4pm. The images were binned by a factor of four for computational efficiency, and aligned by cross-correlation prior to reconstruction. Figure 3a shows the 0° HAADF-STEM image of the sample, and a section through the SIRT reconstruction under the three acquisition scenarios described above (Figure 3b-d).

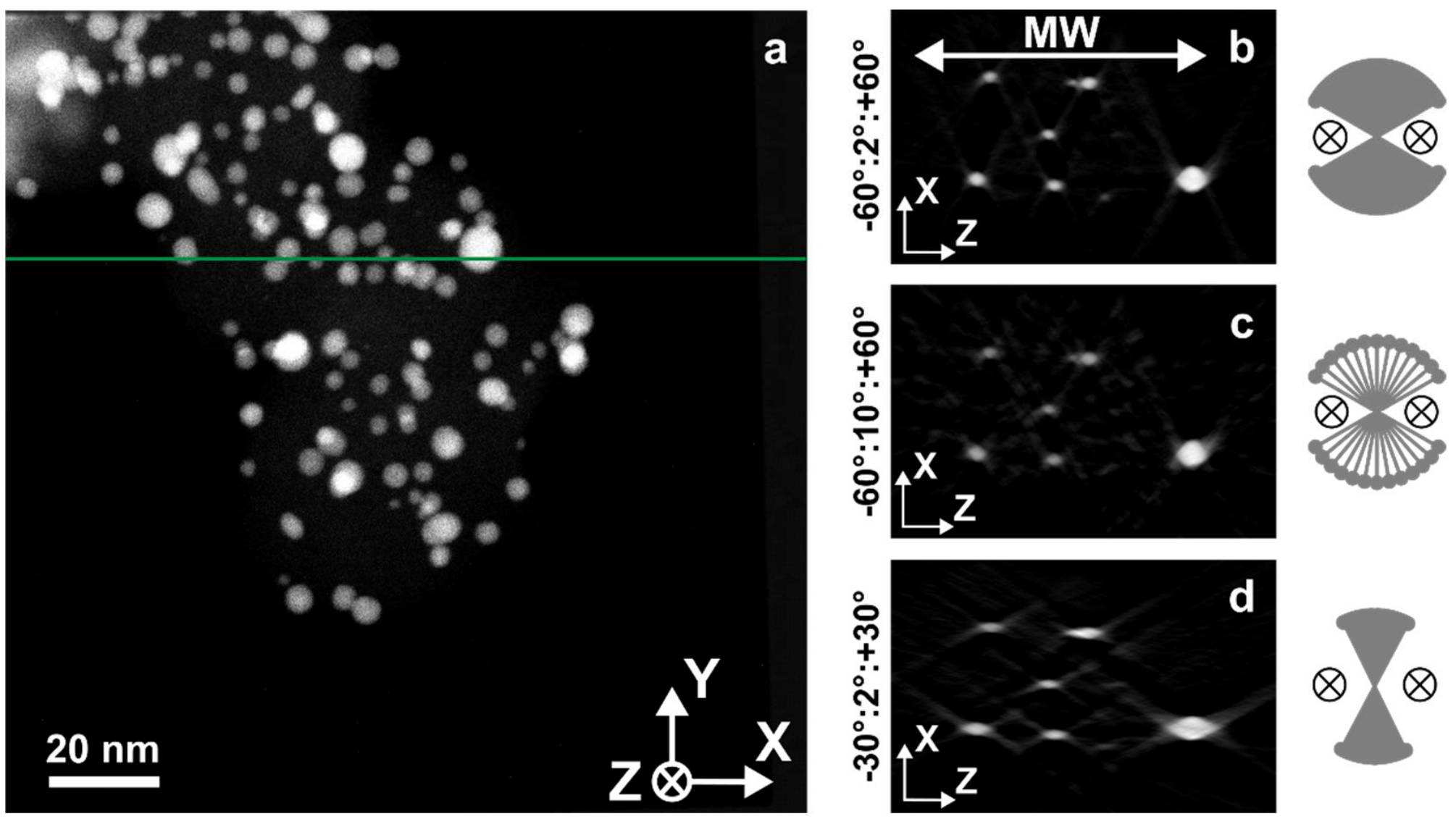

Figure 3. (a) 0° HAADF-STEM image of the Pt/C sample. SIRT reconstruction of a 2D slice (indicated in green in (a)), using [-60°:2°:+60°] (b), [-60°:10°:+60°] (c) and [-30°:2°:+30°] (d) tilt series. The MW is displayed horizontally.

**Benchmarking of various approaches for limited-angle and sparse-view ET**

***Conventional reconstruction approaches: SIRT and CS-TV***

In ET, a 3D volume is reconstructed from a set of 2D projections acquired at different tilt angles. This can be described as an inverse problem:

$$y = Px + n, \tag{1}$$

where $y$ are the experimental projections, $x$ is the object to reconstruct, $P$ is the projection matrix and $n$ is noise. The classical reconstruction method used in ET is the SIRT algorithm which is described as follows:

$$\hat{\boldsymbol{x}} = \underset{x}{argmin}\{\frac{1}{2}\ \|Px - y\|_2^2\}. \tag{2}$$

Throughout this work, ASTRA Toolbox [33] is used for SIRT reconstructions.

A way to improve the reconstruction quality is to impose sparsity of the object in a chosen transformation domain. This translates in a regularization term added to the data fidelity term. In compressed sensing (CS)-based approaches, the regularization term is the *l1*-norm of the object in a chosen transformation domain $L$:

$$\hat{\boldsymbol{x}} = \underset{x}{\operatorname{argmin}} \{ \frac{1}{2} \|Px - y\|_2^2 + \lambda \|Lx\|_1 \}. \tag{3}$$

A common regularization term is the total variation (TV) minimization, where L is the gradient of x, thus promoting piecewise constant intensities and sharp boundaries. The regularization weight $\lambda$ is tailored to each acquisition setting and dataset. Throughout this work, CS-TV reconstructions are performed using the pysap-etomo library [10].

***Deep learning approaches:***

The two DL approaches presented in this work are a supervised U-Net post-reconstruction restoration approach, that we name "SIRT+U-Net", and an unsupervised DIP-based reconstruction approach.

***1/ SIRT + U-Net***

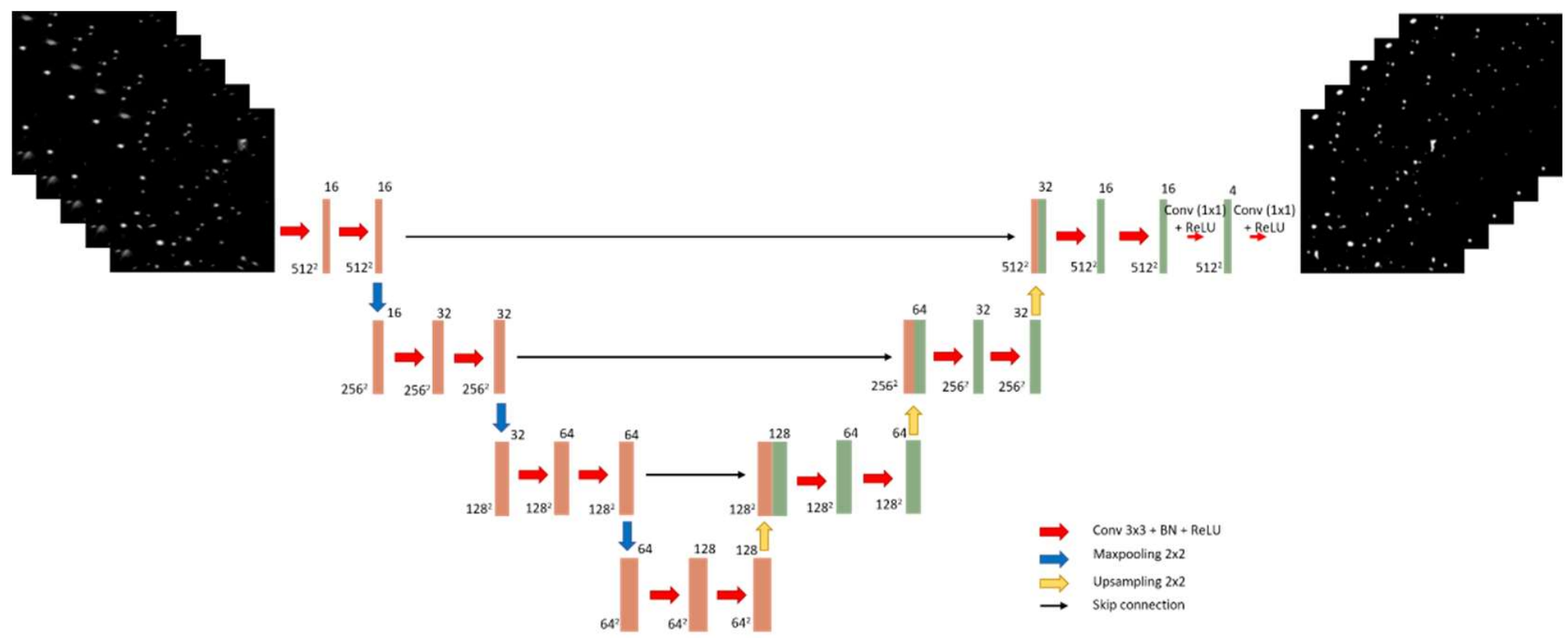

Figure 4: Scheme of the U-Net architecture applied in this work.

In SIRT+U-Net approach, a U-Net model is trained to identify and correct MW artifacts present in the SIRT reconstructions. Since it is a supervised approach, full tilt range data are necessary to generate artifact-free SIRT reconstructions, that serve as reference images for the training the network. Since it is difficult (if not impossible) to generate such data experimentally, the training can be performed using synthetic data similar to the experimental data. In this work, 500 simulated images containing particles of various shapes (mostly spherical) and sizes were generated and used for building tilt series in the following acquisition conditions: [-90°:1°:90°], [-60°:2°:60°], [-60°:10°:60°] and [-30°:2°:30°]. SIRT reconstructions from the first scenario are considered high-quality reference images, while SIRT reconstructions from the other scenarios are considered degraded images. For each acquisition condition, pairs of degraded/high-quality reconstructions are used for the training.

The training procedure can be outlined as follows:

$$\widehat{\boldsymbol{\theta}} = \underset{\theta}{argmin}\{\mathcal{L}(F_\theta(x_{deg}), x_{ref})\}, \quad (4)$$

where $F_\theta$ is the neural network and θ its weights. $\mathcal{L}$ represents the loss function used to measure the error between the high-quality reconstructions and those predicted by the model. The validation set consists of pairs of images generated from 50 simulated images. Once the training phase is completed, the model is employed to restore previously unseen degraded reconstructions:

$$\widehat{\boldsymbol{x}} = \mathrm{F}_{\hat{\theta}}(\mathrm{x}_{\mathrm{deg}}). \quad (5)$$

The neural network used in this work corresponds to a classical U-Net architecture (Figure 4) consisting of 4 blocks, each of them being a succession of 2 convolutional layers with kernel size 3×3 and 16, 32, 64 and 128 channels respectively, followed by a ReLU activation function. Max-pooling is performed between each block in the encoding phase, and 2×2 upsampling is performed in the decoding phase. Skip connections are added between the encoding and decoding phases to use the feature maps from the encoding to the decoding phase. Finally, two single convolutional layers with single kernel and linear activation are added to recover the output image. The training of the model is conducted using the Adam optimizer over 30 epochs, with a batch size of 4 and a learning rate of $3 \times 10^{-4}$. The chosen loss is a mixed $L1$-SSIM (Structural Similarity Index Measure) loss:

$$\mathcal{L}(x_{ref}, x_{out}) = \alpha\left(1 - SSIM(x_{ref}, x_{out})\right) + (1-\alpha)\left|\left|x_{ref} - x_{out}\right|\right|_1, \quad (6)$$

where α is a weighting factor controlling the contribution of each term. In this work, α was set to 0.5.

According to Zhao et al. [34], the mixed $L1$-SSIM loss function outperforms other loss functions, producing superior visual results in image restoration tasks.

The effectiveness of this approach is highly dependent on the quality of the training data and their closeness to the experimental data, in terms of tilt range and increment, but also the level and nature of noise, the resolution, depth of focus and contrast, amongst all the physical conditions that can be simulated but are beyond the scope of this paper.

***2/ DIP approach***

For natural image restoration, the DIP method operates by feeding random noise $z$ into a neural network. The networks' weights are then adjusted to minimize the mean square error (MSE) between the networks' output and the original image intended for restoration:

$$\hat{\theta} = \underset{\theta}{argmin}\{\|F_\theta(z) - I\|_2^2\}, \quad (7)$$

where $I$ denotes the original image to be restored. The optimal restored image is then achieved by employing the trained network on identical input noise, expressed as $\hat{\boldsymbol{I}} = F_{\hat{\theta}}(z)$.

This approach has been modified for x-ray tomography through the incorporation of the Radon transform [32] (Figure 5):

$$\hat{\theta} = \underset{\theta}{argmin}\{\|P(F_\theta(z)) - y\|_2^2\}. \tag{8}$$

In this configuration, the MSE minimization is performed between the experimental sinogram y and the generated reconstructions' sinogram, $P(F_\theta(z))$.

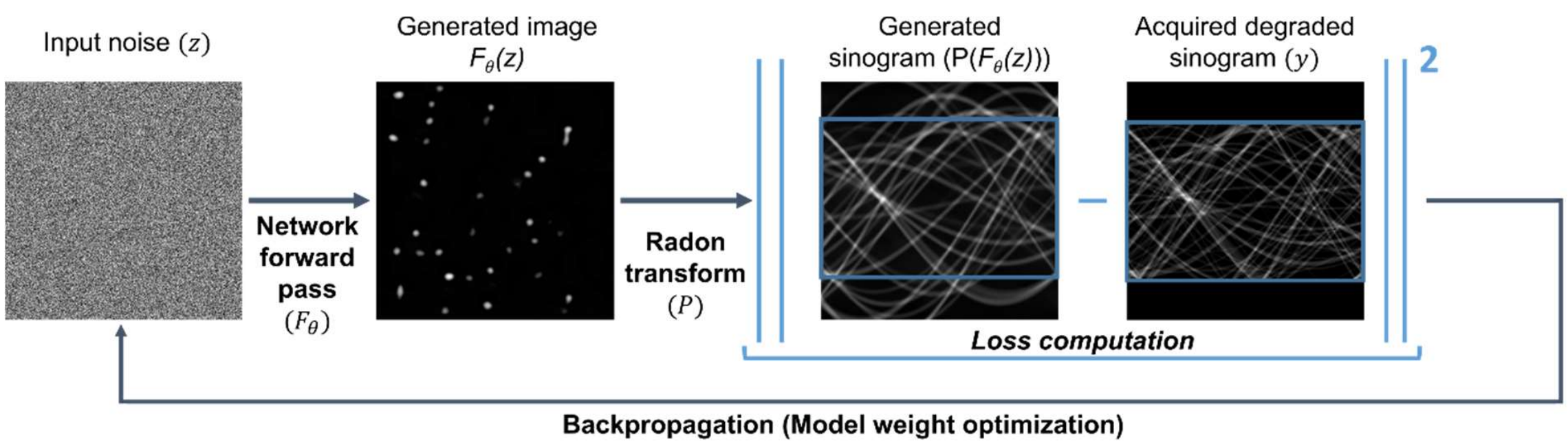

Figure 5: Visual representation of the DIP optimization process for a simulated 2D image. The acquired sinogram contains data in the ±60° angular range.

Here, we use a convolutional U-Net architecture as a neural network (as originally proposed in [31]), which serves as an implicit prior. The number of training iterations is critical when employing the DIP method and is typically determined through empirical testing to ensure convergence and avoid overfitting and undesirable artifacts in the images.

In a similar way to CS-based approaches, a regularization term can be added to the DIP approach:

$$\hat{\theta} = \underset{\theta}{argmin}\{\|PF_\theta(z) - y\|_2^2 + \lambda R\big(F_\theta(z)\big)\}, \tag{9}$$

where $R$ is a regularization term and $\lambda$ is the regularization parameter. In this work, we choose a TV regularization and name the approach "DIP-TV". This DIP-TV approach includes the Radon transform from the Tomosipo library [35] (compatible with Pytorch).

Regarding the simulated data reconstructions, 5000 iterations are performed, selecting the reconstruction with the optimal loss. For the experimental data, 1000 iterations are conducted to minimize noise and overfitting. The network input noise is sampled from a uniform distribution.

All reconstructions, model training and applications were performed on a workstation equipped with an Nvidia A100 GPU (80GB of memory). Training the U-Net on simulated datasets for restoration purposes required approximately three minutes. Reconstruction of a single 512x512 image using DIP-TV took about 25 seconds for 1000 iterations, while reconstruction of the entire experimental volume required roughly 2.5 hours.

### Image quality assessment

The performance of the different approaches was evaluated using both qualitative and quantitative metrics, implemented in IPSDK software and Python.

On simulated data, we evaluated the peak signal-to-noise ratio (PSNR) and SSIM, using full tilt range SIRT reconstructions as references. For quantitative particle analysis, automated Otsu thresholding was applied, followed by 2D watershed separation and connected component labeling. The F1 score was computed from the labeled data to assess the segmentation accuracy relative to the high-quality reference. Additionally, the number of detected particles, along with their equivalent diameter and circularity, were calculated and compared with the corresponding reference values.

The quality of the experimental reconstructions was assessed visually and by comparison with the most sampled scenario ([-60°:2°:60°]. The resulting volumes were then used for 3D particle shape and size analysis. Specifically, 3D segmentation using automated Otsu thresholding was performed, followed by 3D watershed separation and 3D connected component labeling of the NPs. Measurements analogous to those obtained from the simulated data -including the number of reconstructed particles, their equivalent diameter and sphericity- were extracted for the three acquisition conditions.

## Results

### Benchmarking of different approaches on simulated data

The performance of the different approaches is first assessed visually by examining a region of interest in the simulated image (Figure 2). This visual evaluation is summarized in Figure 6, which presents the SIRT, CS-TV, SIRT+U-Net and DIP-TV reconstructions for the three degraded acquisition scenarios ([-60°:2°:60°], [-60°:10°:60°] and [-30°:2°:30°]). As a reference, Figure 6 also includes the SIRT reconstruction obtained from the full tilt-range series.

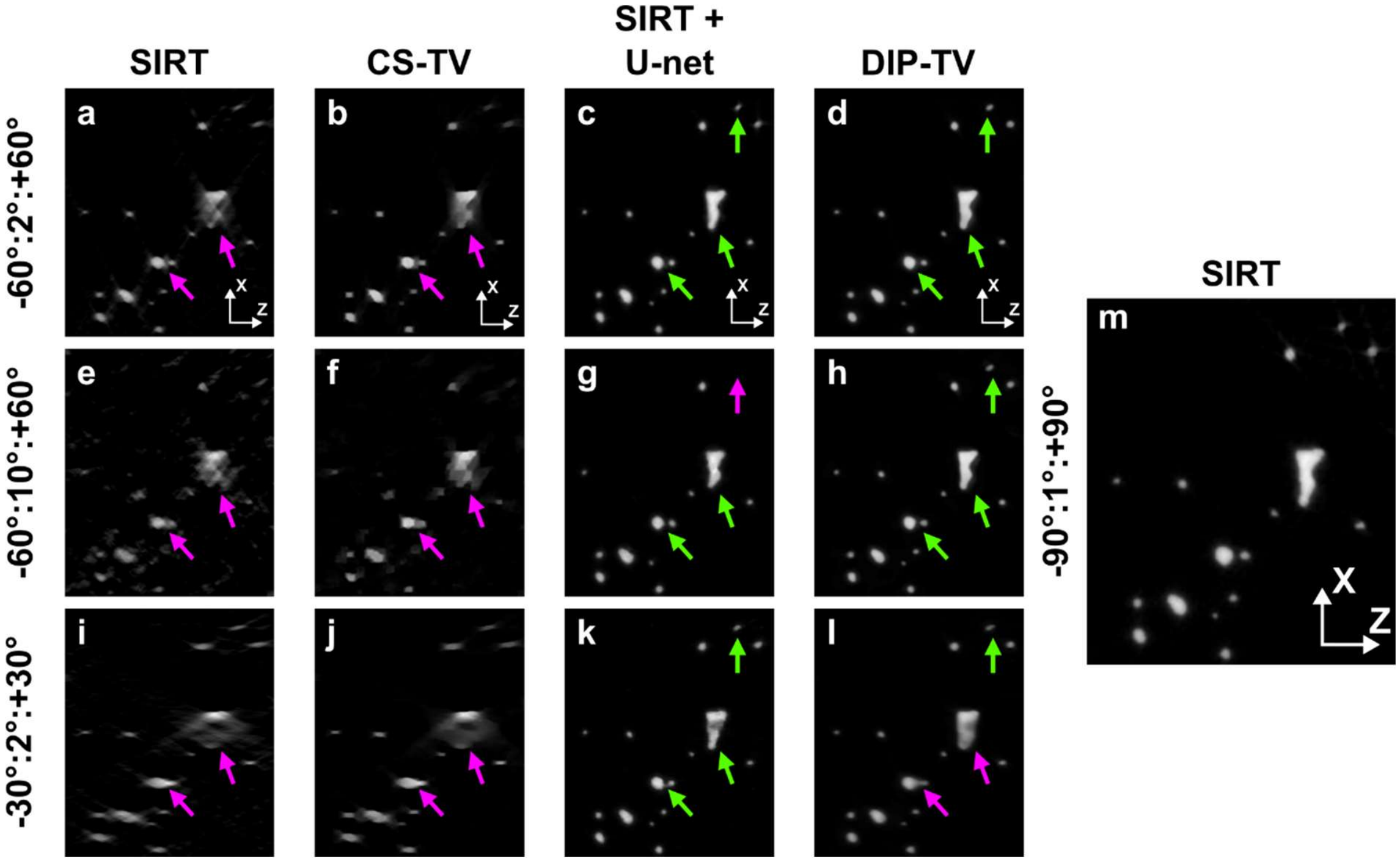

Figure 6. Reconstructions of the simulated image presented in Figure 2, with respectively the SIRT, CS-TV, SIRT + U-Net and DIP-TV approaches, in the [-60°:2°:+60°] (a,b,c,d), [-60°:10°:+60°] (e,f,g,h) and [-30°:2°:+30°] (i,j,k,l) acquisition scenarios and corresponding reference image (m). Green and red arrows indicate respectively well and poorly reconstructed particles.

In the first scenario (Figure 6a-d), CS-TV slightly improves the signal-to-background ratio of the particles compared to SIRT. However, both methods suffer from artifacts induced by the MW. In particular, particles appear elongated, vertical edges are blurred, and neighboring particles along the MW direction are difficult to distinguish reliably. In contrast, both SIRT+U-Net and DIP-TV outperform SIRT and CS-TV, producing reconstructions that are visually very close to the full tilt-range reconstruction.

In the second scenario (Figure 6e-h), SIRT exhibits a noise-like background, making it difficult to reliably distinguish particles from the background. CS-TV again provides a slight improvement in the signal-to background ratio, but, as in the first scenario, both methods are strongly affected by the combined MW and sparse-view conditions. Small particles are not recovered, the vertical edge is lost, and adjacent particles are merged. SIRT+U-Net outperforms the classical approaches but fails to recover small isolated particles. This limitation is partly due to the fact that the model is trained on SIRT reconstructions, in which such particles are not visible under sparse-view conditions and therefore cannot be retrieved. In contrast, DIP-TV successfully retrieves all particles with reliable shapes and intensities.

For the extreme case (Figure 6i-l), the reconstructions obtained using classical approaches are severely degraded and largely unexploitable. SIRT+U-Net significantly suppresses elongation artifacts and successfully recovers all particles, exhibiting higher performance than in the sparse-view scenario. DIP-TV also retrieves all isolated particles, but does not adequately recover neighboring particles and may require a higher number of iterations and finer parameter tuning. Nevertheless, the results are remarkable for an *un*supervised strategy.

The conclusions from this visual assessment are also supported by the image quality metrics and quantitative measurements computed on the validation dataset (50 images) in Figure 7, using the full tilt-range SIRT reconstructions as references. The SSIM and PSNR metrics confirm the superiority of both DL approaches across all three scenarios, with DIP-TV performing slightly better in the sparse-view scenario and SIRT+U-Net in the highly limited-angle case.

The F1 score and the number of detected particles, obtained after binarization and labeling, further demonstrate that the DL approaches can reliably segment and detect the simulated particles. Notably, SIRT overestimates the number of particles due to noise in the reconstructions. Finally, the mean particle area and circularity confirm that the DL approaches provide reliable quantitative measurements in all three scenarios.

**SSIM**

| | -60°:2°:+60° | -60°:10°:+60° | -30°:2°:+30° |
|---|---|---|---|
| SIRT | 0.850 | 0.597 | 0.674 |
| CS-TV | 0.904 | 0.747 | 0.762 |
| SIRT + U-Net | **0.974** | 0.905 | **0.938** |
| DIP-TV | 0.949 | **0.929** | 0.906 |

**PSNR**

| | -60°:2°:+60° | -60°:10°:+60° | -30°:2°:+30° |
|---|---|---|---|
| SIRT | 31.6 | 27.4 | 26.3 |
| CS-TV | 32.6 | 29.4 | 27.1 |
| SIRT + U-Net | **40.3** | 33.1 | **34.3** |
| DIP-TV | 37.1 | **35.6** | 32.5 |

**F1 Score**

| | -60°:2°:+60° | -60°:10°:+60° | -30°:2°:+30° |
|---|---|---|---|
| SIRT | 0.863 | 0.709 | 0.630 |
| CS-TV | 0.874 | 0.819 | 0.672 |
| SIRT + U-Net | **0.956** | 0.885 | **0.906** |
| DIP-TV | 0.919 | **0.904** | 0.829 |

**Number of particles** (*GT: 2843*)

| | -60°:2°:+60° | -60°:10°:+60° | -30°:2°:+30° |
|---|---|---|---|
| SIRT | 2823 | 3469 | 3033 |
| CS-TV | 2679 | 2323 | 2999 |
| SIRT + U-Net | **2827** | 2355 | 2714 |
| DIP-TV | 2803 | **2694** | **2768** |

**Mean area in pixels** (*GT: 60.7*)

| | -60°:2°:+60° | -60°:10°:+60° | -30°:2°:+30° |
|---|---|---|---|
| SIRT | 70.9 | 72.0 | 103.5 |
| CS-TV | 69.0 | 78.0 | 96.3 |
| SIRT + U-Net | **63.2** | 71.4 | **58.0** |
| DIP-TV | **63.3** | **64.8** | 73.1 |

**Mean circularity** (*GT: 0.804*)

| | -60°:2°:+60° | -60°:10°:+60° | -30°:2°:+30° |
|---|---|---|---|
| SIRT | 0.764 | 0.687 | 0.653 |
| CS-TV | 0.812 | 0.796 | 0.737 |
| SIRT + U-Net | 0.803 | 0.799 | 0.791 |
| DIP-TV | **0.804** | **0.802** | **0.796** |

Figure 7. Tables with the computation of image reconstruction quality metrics (SSIM, PSNR and F1 score) (top row) and particles shape and sizes analysis (number, mean area and mean circularity) (bottom row) on the reconstructions from the simulated data sets for each reconstructions approaches and acquisition scenarios.

Figure 8 presents histograms of particle circularity for the three scenarios and the four reconstruction approaches, showing the deviation of the measurements from the reference values. These results demonstrate that the DL methods are consistently aligned with the reference in all scenarios, and that the unsupervised approach performs comparably to the supervised one while requiring no training dataset, making it applicable to a broader range of experiments and sample types.

For STEM-HAADF tomography in materials science and for BF-TEM in biology, the most widely used acquisition scheme corresponds to the [-60°:2°:+60°] configuration. In these conditions, the quality of SIRT+U-Net and DIP-TV reconstructions, together with the associated quantitative measurements, suggests that these approaches could substantially improve the quality of reconstructions obtained from standard acquisitions, thereby facilitating segmentation and enhancing the reliability of quantitative analysis.

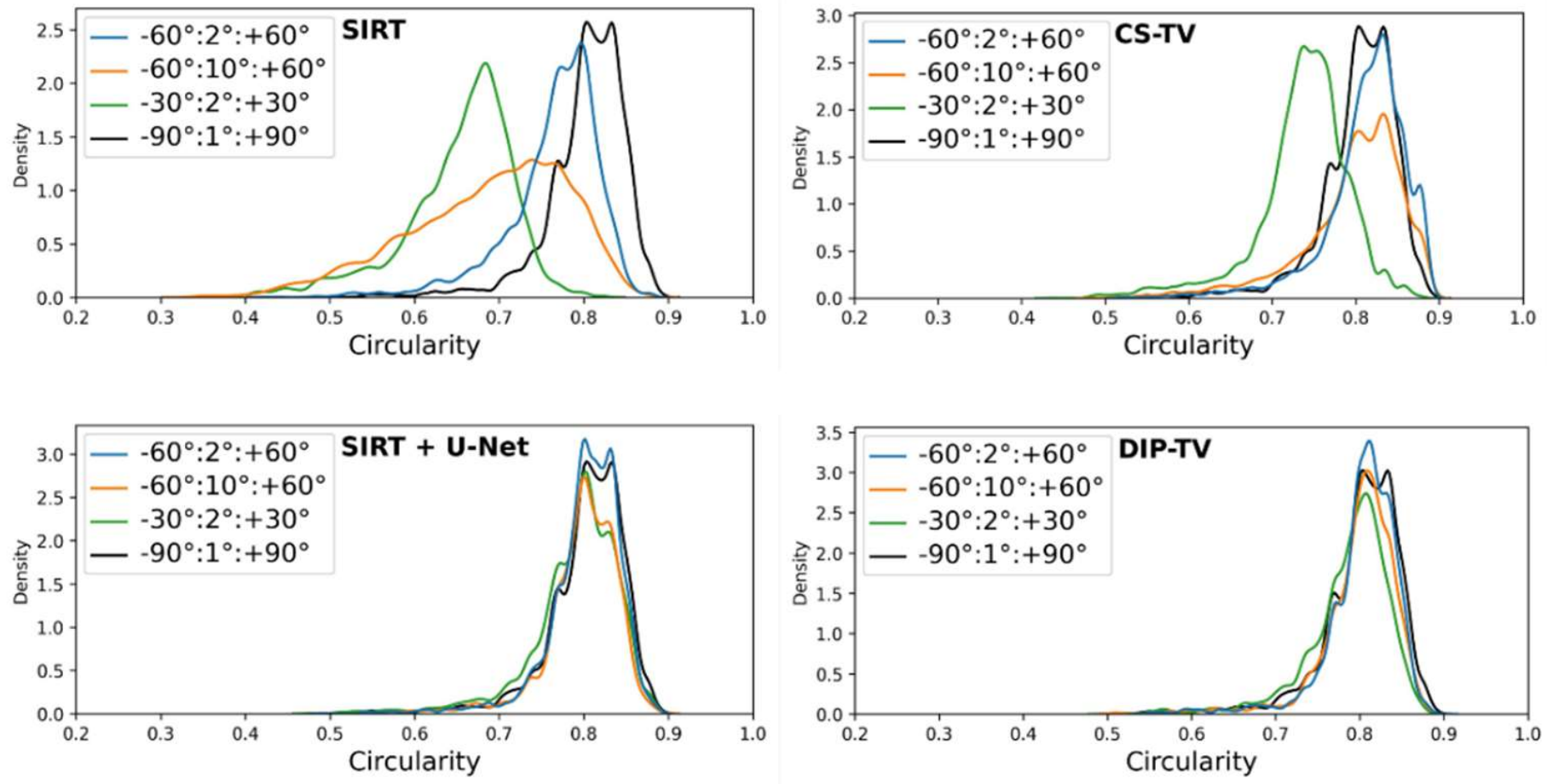

Figure 8. Graphs representing the distribution of the circularity of the simulated particles for each reconstruction approach (SIRT, CS-TV, SIRT+U-Net and DIP-TV) and each acquisition scenario ([-60°:2°:+60°] (blue), [-60°:10°:+60°] (orange) and [-30°:2°:+30°] (green) and the reference acquisition scenario distribution [-90°:2°:+90°] (black).

**3D morphological characterization of platinum nanoparticles by STEM-HAADF tomography**

In this section, we benchmark the different reconstruction approaches for the 3D morphological study of Pt NPs using STEM-HAADF tomography. The experimental tilt series was acquired using the conventional acquisition scheme ([-60:2°:+60°]). As no experimental ground truth is available for direct comparison, undersampled reconstructions (scenario 2 and 3) are compared against the highly sampled case.

Figure 9 presents the 3D renderings of the reconstructed volumes for the three acquisition scenarios using SIRT, CS-TV, SIRT+U-Net and DIP-TV. Additionally, in the Supporting Information, we incorporate slices through the labelled reconstructions for each scenario and methodology. In the first and second scenarios (Figure 9a-h), CS-TV and the DL approaches produce visually similar results, whereas in the third scenario (Figure 9i-l) only DL approaches remain visually consistent. This observation is further supported by Figure 10, which shows selected slices through the reconstructions with neighboring particles, allowing us to evaluate which method can detect and separate them accurately. The MW direction in both figures is oriented horizontally.

In the first scenario (Figure 9a-d), and as expected from the simulation results, CS-TV provides reconstructions with a higher signal-to-background ratio than SIRT. However, for both approaches, neighboring particles along the MW direction are difficult to separate as observed in Figure 10a-b, and small particles exhibit lower intensities than larger ones, making the thresholding step challenging. In contrast, the SIRT+U-Net and DIP-TV reconstructions show better-resolved particles that can be more easily separated. This improvement is reflected in the particle counts in Figure 11: DL approaches detect more particles than the conventional methods when using the same thresholding and labeling procedure. This effect is more pronounced for the experimental data than for the simulated datasets, likely due to the higher prevalence of neighboring particles in the experimental sample. In the sparse-view scenario (Figure 9e-h), neighboring particles and small isolated particles are poorly recovered using SIRT and CS-TV, whereas SIRT+U-Net and DIP-TV exhibit clear improvements as observed in Figure 10g-h. In particular, the signal-to-background ratio is substantially enhanced, and small particles are reliably recovered with both approaches. Nevertheless, DIP-TV outperforms SIRT+U-Net in the thresholding and labeling of neighboring particles as supported by the quantitative measurements in Figure 11, consistent with the simulation results. This trend is also reflected in the mean equivalent diameter and sphericity measurements observed in Figure 12.

In the highly limited-angle case (Figure 9i-l), SIRT and CS-TV yield unreliable reconstructions, whereas SIRT+U-Net and DIP-TV successfully recover most of the particles. Among all methods, SIRT+U-Net performs particularly well at separating neighboring particles, as shown in Figure 10k-l, where DIP-TV shows more difficulty separating them. However, this difference is less pronounced overall: in Figure 11, both DL methods yield the same particle count, and considering that SIRT+U-Net is supervised, DIP, as unsupervised, can be advantageous in terms of generalization across specimen and acquisition conditions. This is also supported by the mean equivalent diameter values in Figure 11. In this highly limited-angle scenario, SIRT+U-Net gives the smallest mean equivalent diameter (4.65 nm), closely followed by DIP (4.80 nm), i.e., a difference of only 0.15 nm. In general, the mean absolute difference between the two DL methods is small (0.22 nm on average across scenarios), while the discrepancy between the classical and DL methods is larger (0.50 nm on average). In addition, it becomes extreme in the most limited-angle case (0.82-1.12nm), i.e, about 5-7 times larger than the difference between DL strategies. Thus, we conclude that a properly fine-tuned DIP can achieve results comparable to SIRT+U-Net, while offering the practical advantages of an unsupervised approach.

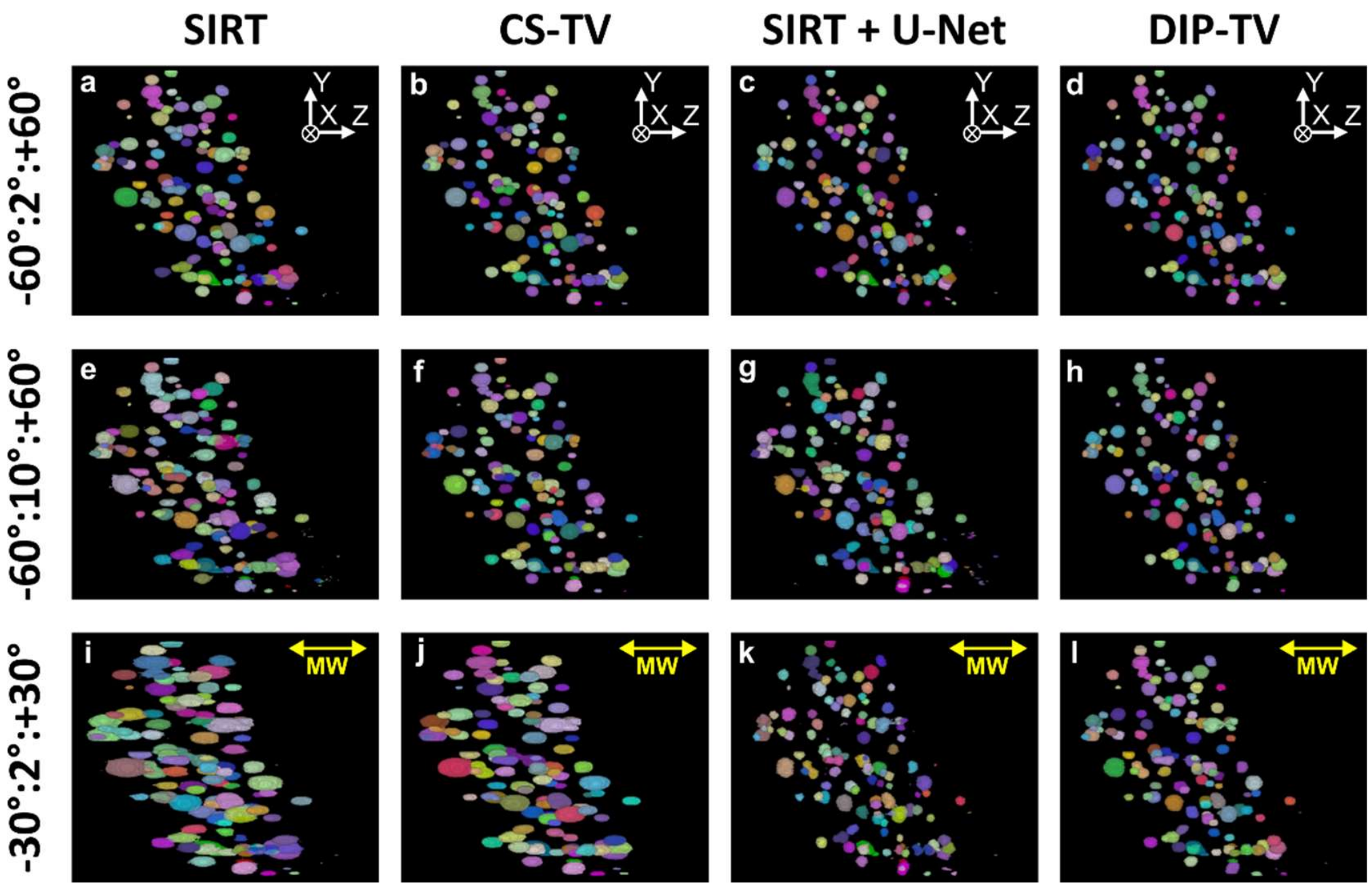

Figure 9. 3D rendering of the reconstructions of the Pt NPs with the four reconstruction techniques used: SIRT, CS-TV, SIRT+U-Net and DIP-TV (in columns), and the three acquisition scenarios (a-d): [-60°:2°:+60°], (e-h) [-60°:10°:+60°] and (i-l) [-30°:2°:+30°].

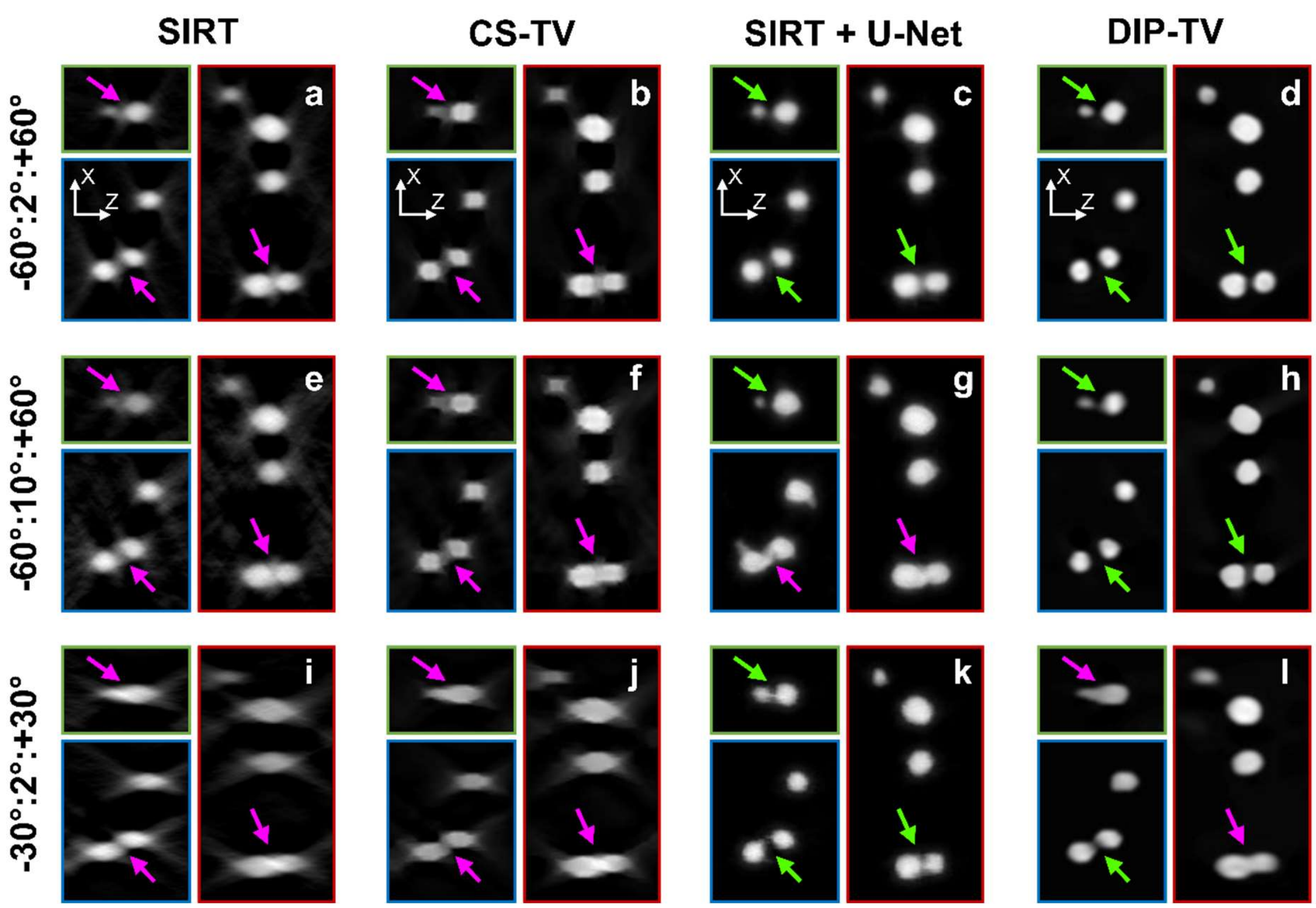

Figure 10. Selected slices through the experimental volumes reconstructed with respectively the SIRT, CS-TV, SIRT + U-Net and DIP-TV approaches, in the (a-d) [-60°:2°:+60°], (e-h) [-60°:10°:+60°] and (i-l) [-30°:2°:+30°] acquisition scenarios.

| | Number of particles | | | Mean equivalent diameter (nm) | | | Mean sphericity | | |
|---|---|---|---|---|---|---|---|---|---|
| | -60°:2°:+60° | -60°:10°:+60° | -30°:2°:+30° | -60°:2°:+60° | -60°:10°:+60° | -30°:2°:+30° | -60°:2°:+60° | -60°:10°:+60° | -30°:2°:+30° |
| **SIRT** | 116 | 112 | 107 | 4.94 | 5.26 | 5.77 | 0.859 | 0.786 | 0.714 |
| **CS-TV** | 115 | 112 | 110 | 4.78 | 4.87 | 5.61 | 0.875 | 0.841 | 0.769 |
| **SIRT + U-Net** | **120** | **119** | **116** | 4.70 | 5.00 | **4.65** | **0.889** | 0.815 | 0.821 |
| **DIP-TV** | 119 | 118 | **116** | **4.59** | **4.60** | 4.80 | 0.885 | **0.875** | **0.839** |

Figure 11. Tables with the computation of particles shape and size analysis (number, mean equivalent diameter and mean sphericity) on the reconstructions from the experimental data for each reconstruction approach.

Moreover, Figure 11 reports the mean sphericity for each reconstructed volume. SIRT is highly sensitive to the tilt range and angular increment, whereas CS-TV is more stable under sparse-view conditions than under limited-angle conditions. In the sparse-view scenario, DIP-TV better preserves particle shape (higher sphericity), while in the highly limited-angle scenario ([-30°:2°:+30°]), SIRT+U-Net yields slightly higher sphericity. Figure 12 further supports these results. The classical methods show a pronounced scenario-dependent shift and broadening of the sphericity and equivalent diameter distributions. In contrast, DL approaches yield tighter, more overlapping distributions across acquisition scenarios, indicating more robust and consistent shape and size estimates.

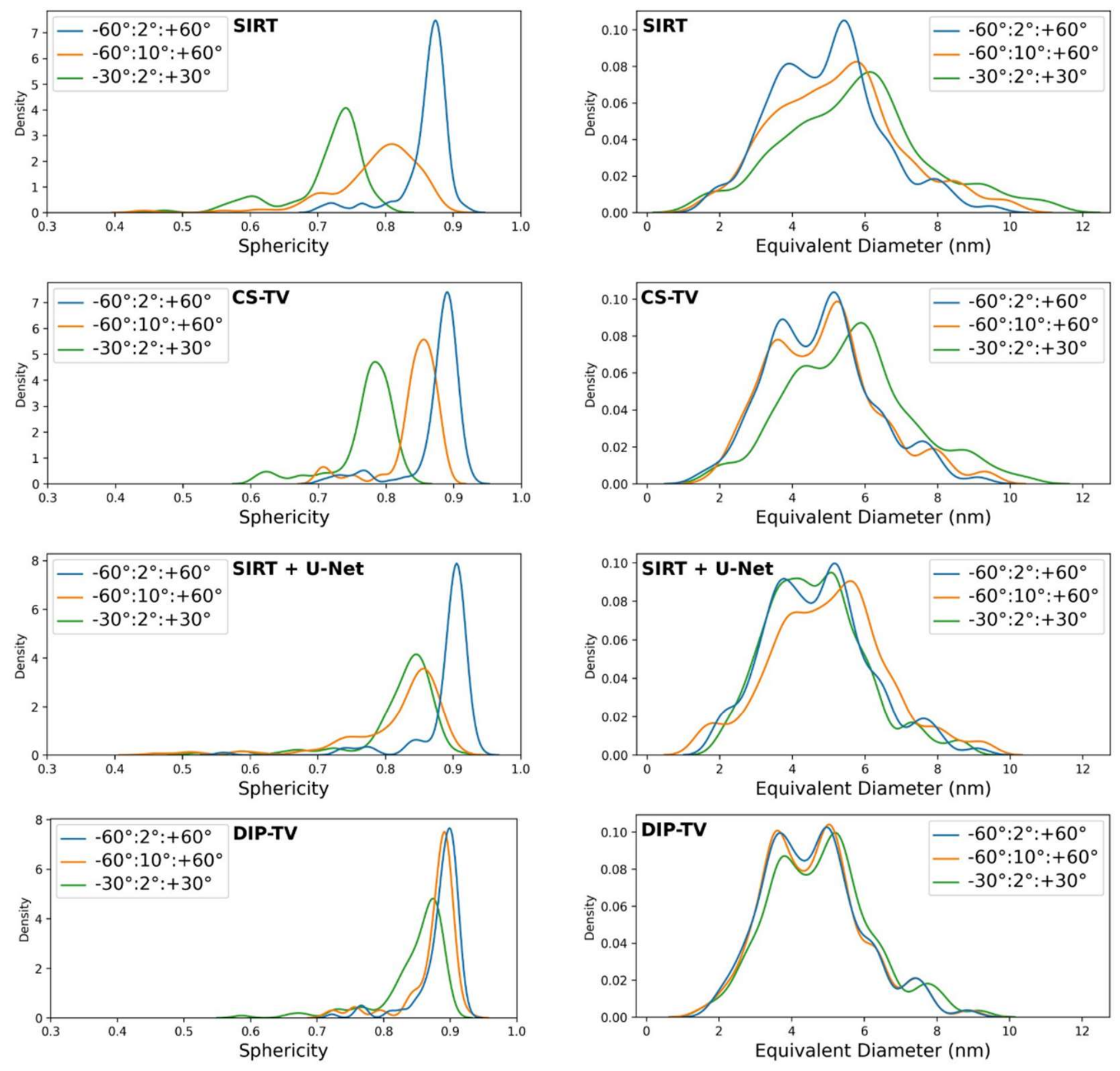

Figure 12. Graphs representing the distribution of the sphericity (left column) and equivalent diameter (right column) of the Pt NPs particles for each reconstruction approach (SIRT, CS-TV, SIRT + U-Net and DIP-TV) and each acquisition scenario ([-60°:2°:+60°] (blue), [-60°:10°:+60°] (orange) and [-30°:2°:+30°] (green).

Overall, DIP-TV demonstrates very stable performances, with only minor variations in the quantitative measurements as the acquisition scenario becomes more challenging. The supervised approach could likely be further improved by generating more realistic simulations that incorporate the image formation process and noise levels present in the experimental data. However, such training remains dependent on the specific tilt range and angular increment used for the experimental acquisitions.

## I. Discussion

The DL-based methods investigated in this work consistently outperform traditional SIRT and CS-TV reconstruction techniques across all examined scenarios. In the standard acquisition scenario, both DL methods yield comparable results. The slightly higher performance of SIRT+U-Net in the ±30° case reflects task-specific supervised training, whereas DIP achieves similar performance -and in the sparse-view case, superior shape preservation- without requiring any specimen-specific training dataset, making it more broadly applicable and potentially more generalizable across acquisition scenarios and specimens.

Both DL approaches have been implemented in a Python library and released as open-source (https://github.com/CEA-MetroCarac/DL_etomo). As discussed previously, the performance of the supervised approach strongly depends on the quality and representativeness of the training datasets. While training and applying the model to similar datasets –such as simulated NPs– is relatively straightforward, discrepancies between the training data and the experimental data can lead to suboptimal restorations. In tomographic reconstruction, tools such as the ASTRA Toolbox enable the simulation of projections with sparse-view or MW conditions. However, ET datasets can exhibit substantial variability, and supervised approaches may not be optimal in scenarios where the samples are too complex to be realistically simulated.

In contrast, the DIP approach offers greater generalization potential and can be applied to a wide range of datasets without prior training. However, DIP-based methods still present some challenges, such as determining model convergence and selecting an appropriate number of iterations. These parameters lack universal selection criteria and typically require case-by-case tuning through multiple trials. In addition, the choice of the TV regularization weight must be carefully adjusted for each new reconstruction, similarly to the CS-based approaches. Nevertheless, despite these practical considerations, the absence of training data requirements remains a decisive advantage for DIP in realistic ET experiments.

The impact of noise has not been investigated in this study, and the proposed methods may therefore not be directly applicable when a significant level of noise is present in the acquired projections. Addressing low-dose scenarios would require investigating alternative regularization terms, joint multi-channel reconstruction strategies, or other learning-based algorithms for tomographic reconstructions. For instance, unrolled neural networks –which involve unrolling the iterations of a reconstruction algorithm as sequential layers of a neural network– have demonstrated increased robustness and adaptability to previously unseen data. In particular, learned primal-dual networks have attracted substantial attention in the fields of CT and MRI reconstruction [36-38].

Alternative approaches, such as plug-and-play (PnP) methods [39-40], provide a reliable option by integrating a pretrained denoiser within an iterative tomographic reconstruction framework, similarly to CS-based approaches. A key advantage of PnP methods over unrolled techniques lies in their flexibility, as they do not require retraining when the acquisition setup varies. Finally, unsupervised methods based on implicit neural representations, applied in a manner similar to the DIP framework, have demonstrated excellent performance in CT reconstruction [41-42] and show strong potential, particularly for spectroscopic ET and cryo-ET.

## II. Conclusion

In this study, we investigated two distinct DL methodologies for the 3D quantitative analysis of Pt NPs using HAADF-STEM tomography. The first approach employs a supervised U-Net restoration model, while the second relies on an unsupervised DIP-based strategy. Both methods substantially outperform reconstruction techniques commonly used in ET. Under a standard acquisition scheme (e.g. [-60°:2°:+60°]), both approaches produce high-quality reconstructed volumes, thereby facilitating subsequent segmentation. For sparse-view and limited-angle acquisitions, both approaches surpass conventional methods. The supervised U-Net restoration model performs particularly well in experiments with extremely limited tilt ranges ([-30°:2°:+30°]), although this advantage is expected since we trained it on a dataset specifically constructed to match the experimental sample, which in this study was easily simulated. In addition, the unsupervised DIP-based approach excels under sparse-view acquisitions ([-60°:10°:+60°]), while still producing remarkable reconstructions under limited-angle conditions. This makes DIP particularly useful when representative training data cannot be generated. Overall, this unsupervised approach shows considerable promise for improving the resolution of ET reconstructions across a wide range of samples, experimental setups, and acquisition modalities, including EELS, EDX and ptychography. The effectiveness of DIP-based approaches for STEM-EDX tomography will be investigated in a separate study [43], with a particular focus on low-dose acquisitions and dose-reduction strategies.

# Acknowledgments:

This work, carried out on the Platform for Nanocharacterisation (PFNC), was supported by the "Recherche Technologique de Base" program of the French ministry of research. The authors also acknowledge the French National Research Agency through the PEPR DIADEM, David Cullen (ORNL) for providing the samples, Guillaume Biagi for his implication in the U-Net implementation, and Nicola Vigano for insightful discussions on the DIP method.